\RequirePackage{fix-cm}
\documentclass[column]{article}  
\RequirePackage{graphicx}
\usepackage[utf8]{inputenc}
\usepackage[T1]{fontenc}
\usepackage{graphicx}
\usepackage{grffile}
\usepackage{longtable}
\usepackage{wrapfig}
\usepackage{rotating}
\usepackage[normalem]{ulem}
\usepackage{amsmath}
\usepackage{textcomp}
\usepackage{amssymb}
\usepackage{capt-of}
\usepackage{hyperref}
\usepackage{listingsutf8}
\usepackage{subcaption}
\usepackage{xcolor}
\usepackage{placeins}
\usepackage{tikz}
\usepackage{authblk}
\graphicspath{{./}}

\newcommand{\nuc}[2]{\(^{#1}\)#2}
\newcommand{\gr}{{$\gamma$ ray}}
\newcommand{\gdr}{{$\gamma$-ray}}
\newcommand{\grs}{{$\gamma$ rays}}
\newcommand{\gdrs}{{$\gamma$-rays}}
\newcommand{\grt}{{$\gamma$-ray tracking}}
\newcommand{\atac}{Ata\k{c}}
\newcommand{\senyigit}{\k{S}enyi\^{g}it}

\newcommand{\lalovic}{Lalovi{\'c}}

\begin{document}

\title{AGATA: Performance of $\gamma$-ray tracking and associated
  algorithms}

\author[1]{F.C.L. Crespi} 
\author[2]{J. Ljungvall\footnote{Corresponding author: joa.ljungvall@ijclab.in2p3.fr}}
\author[2]{A. Lopez-Martens}
\author[3]{C. Michelagnoli}

\affil[1]{Dipartimento di Fisica, Universit\`a degli Studi 
  di Milano, I-20133 Milano, Italy\\
  and\\
  INFN Sezione di Milano, I-20133 Milano, Italy}
\affil[2]{  Universit\'e Paris-Saclay, CNRS/IN2P3,
  IJCLab, 91405 Orsay, France}
\affil[3]{
  Institut Laue-Langevin,
  71 Avenue des Martyrs,
  38042 Grenoble, France}

\date{Received: date / Accepted: date}

\maketitle

\begin{abstract}
AGATA is a modern $\gamma$-ray spectrometer for in-beam nuclear
structure studies, based on $\gamma$-ray tracking. Since more than a
decade, it has been operated performing experimental physics campaigns
in different international laboratories (LNL, GSI, GANIL). This paper
reviews the obtained results concerning the performances of
$\gamma$-ray tracking in AGATA and associated algorithms. We discuss
$\gamma$-ray tracking and algorithms developed for AGATA. Then, we
present performance results in terms of efficiency and peak-to-total
for AGATA. The importance of the high effective angular resolution of
$\gamma$-ray tracking arrays is emphasised, e.g. with respect to
Doppler correction. Finally, we briefly touch upon the subject of
$\gamma$-ray imaging and its connection to $\gamma$-ray tracking.
\\
\textbf{Keywords~} AGATA $--$ Gamma-ray spectroscopy $--$ Gamma-ray tracking
    $--$ Gamma-ray imaging
\end{abstract}
\section{Introduction}
\label{sec:intro}

The Advanced GAmma Tracking Array (AGATA)~\cite{Akkoyun201226} is the
European state-of-the-art high-resolution $\gamma$-ray spectrometer. A
similar project in the US is called GRETINA/GRETA
\cite{LEE1999195,LEE2004255}. In past decades, developments in the
technology of semiconductor (HPGe) detectors resulted in,
systematically, to significant advancements in nuclear science. In
fact, since the first generation of large arrays of HPGe detectors
($\sim$1981) became available, high-resolution spectroscopic
experiments played a primary role in illuminating fundamental aspects
of nuclear structure. With the second generation arrays, based on
Compton-suppressed HPGe detectors (e.g. GAMMASPHERE, EUROBALL), this
technology reached a saturation point in terms of experimental
sensitivity for in-beam high-resolution $\gamma$ spectroscopy studies
(see e.g. Lee at al.~\cite{Lee_2003} and Eberth et al.
\cite{EBERTH2008283}). It was then realised that the possibility to
overcome this technological limit was given by being able to produce
new HPGe detectors featuring millimetric position resolution, obtained
through the physical segmentation of the detector electrode and the
analysis of the shapes of the electric signals. This kind of segmented
HPGe detectors allows, in fact, to reconstruct the path of a $\gamma$
ray that interacted in their active volume. In terms of experimental
sensitivity advantages, the $\gamma$-ray tracking technology permits,
for example, to maximise the detection efficiency (due to the
elimination of the Compton shields), and to recover the energy
resolution degradation due to the Doppler effect. This while retaining
a good suppression of radiation not coming from the target position.
These features fit specific needs of in-beam high-resolution $\gamma$
spectroscopy experiments with radioactive beams, that are at the
forefront of present nuclear structure research.

The AGATA array is designed to be composed of 180 36-fold segmented
HPGe detectors, grouped in 60 triple clusters (final configuration),
resulting in a coverage of 80\% of 4$\pi$ of solid angle around the
target position. In the present phase of the project a coverage of
1$\pi$ solid angle has been reached, corresponding to 15 AGATA Triple
Clusters (ATCs). Recently a Memorandum of Understanding was signed
between the different partner institutes for a phase 2 of the project,
bringing the array up to a solid coverage of $3\pi$ in 2030. In this
manuscript we review the performance of the $\gamma$-ray tracking in
AGATA, referring also to specific opportunities of accessing novel
experimental physics information as compared with previous generation
HPGe arrays. In the initial phase of the AGATA project the
performances of $\gamma$-ray tracking were extensively studied via
Geant4 simulations \cite{Farnea2010331}.

Presently, after about 12 years of operation in experimental physics
campaigns in different international laboratories (LNL, GSI, GANIL),
these performances can be highlighted with reference to many
experimental results, as it will be shown in the following. For a
recent review of the scientific output and perspectives of AGATA, see
e.g.~\cite{BRACCO2021103887,KortenW2020Powt}. 
\section{General considerations for $\gamma$-ray tracking}
\label{sec:gsfgrt}

Since Compton scattering is the dominant interaction mechanism of
photons in Ge for energies ranging from 150 keV to 10 MeV, tracking
algorithms are mainly based on the properties of the Compton
interaction process.
  
In particular, they rely on the following relationship
between incident $E_{i-1}$ and scattered $E_{i}$ energies and
scattering angle $\theta_i$ (assuming that the electrons of the Ge
detectors are at rest):
\begin{center}
\begin{equation}
cos(\theta_i) = 1-m_ec^2\left(\frac{1}{E_{i}}-\frac{1}{E_{i-1}}\right)
\label{compt}
\end{equation}
\end{center}

Most tracking algorithms attempt to reconstruct the tracks of photons,
which have been fully absorbed in the Ge detectors (there is the
notable exception of the TANGO algorithm~\cite{TASHENOV2010592}, which
also identifies Compton escape events, see section~\ref{sec:ogrta}).
There are two categories of tracking algorithms: forward-tracking
algorithms~\cite{Schmid199969}, which start from the known position of
the source and reconstruct the track of photons as they interact in
the detector and back-tracking algorithms~\cite{vanderMarel1999538},
which start from the potential photoelectric interaction point and
reconstruct the track backwards to the source. Forward-tracking
algorithms have been demonstrated to be more efficient than
back-tracking algorithms~\cite{LopezMartens2004454} and are therefore
used both at AGATA~\cite{Akkoyun201226} and GRETINA
\cite{Paschalis201344}.

Three algorithms have been used to track AGATA data: The Mars Gamma
Tracking algorithm~\cite{BAZZACCO2004248} (MGT) and the Orsay Forward
Tracking algorithm~\cite{LopezMartens2004454} (OFT), which are both
implemented into the AGATA data acquisition software, and the
Gretina Tracking algorithm~\cite{LAURITSEN201646,Korichi2019}
(henceforth called GT). These algorithms are composed of two parts:
the first one consists in defining a pool of clusters of interaction
points in 3-dimensional space and the second part consists in finding
the best sequence of interaction points for each cluster and keeping
only those sequences, whose figure of merit lies above a given
threshold.

The clusterisation of points is typically performed on the basis of
the angular separation between points. The free parameter is then the
maximum opening angle, as is the case in GT. For OFT and MGT, the
maximum angle depends on the number of interaction points detected in
each event and is therefore not a tuneable parameter. Other more
sophisticated clusterisation algorithms have been developed in the
framework of AGATA, such as the fuzzy C-means algorithm
\cite{Suliman2010} (see section~\ref{sec:ogrta}) or the Deterministic
Annealing Filter~\cite{DIDIERJEAN2010188}).

\begin{figure}[tbh]
\begin{center}
\includegraphics[width=\columnwidth]{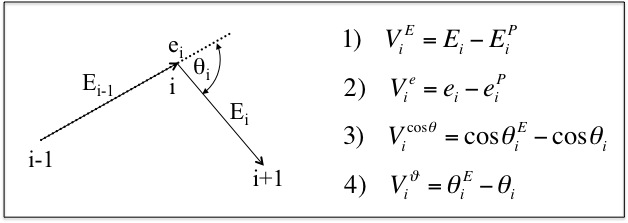}
\end{center}
\caption{Example of a Compton scattering vertex, in which an incident
  photon of energy $E_{i-1}$, deposits the energy $e_i$ at vertex $i$
  and is scattered at an angle $\theta_i$ with energy $E_{i}$. The
  superscript $P$ indicates scattered and deposited energies obtained
  through measured interaction-point (or source) positions and the
  superscript $E$ indicates angles or cosines calculated from the
  measured deposited energies. }
\label{vertex}
\end{figure}

The evaluation of clusters containing more than one interaction point
then proceeds as follows: starting from the known position of the
source, and assuming that the total deposited energy in a cluster
corresponds to the energy of an incident photon, the figure of merit
of every possible sequence of interaction points within the cluster is
computed by evaluating the goodness of each vertex in the sequence by
comparing the measured energies and angles with the corresponding
quantities obtained via equation~\ref{compt}. As shown in
Fig.~\ref{vertex} this comparison can be performed in different ways.

GT uses the last comparator $V_i^{\theta}$ of Figure~\ref{vertex},
while OFT and MGT use the first comparator $V_i^E$ with an added
weight, which accounts for range and interaction-process probabilities
in going from interaction-point $i-1$ to $i$ and $i+1$. A weighted
square sum is of these vertices is used by MGT while OFT uses a
likelihood like formulation. For OFT, the comparator $V_i^E$ is also
weighted by the associated experimental uncertainties in the deposited
energies and interaction positions.

The evaluation of single-interaction-point \lq\lq clusters\rq\rq\ is
an important part of all tracking algorithms since the efficiency loss
when it is not included is very large for low energy events, and
non-negligible at higher energies. As a consequence of how the Pulse
Shape Analysis (PSA) algorithm (Adaptive Grid
Search~\cite{ventrurelli2004}) identifies interaction points in AGATA,
$\sim$20$\%$ of 1.4 MeV total-absorption events in AGATA detectors are
found in single interaction points. This is at variance with the
situation at GRETINA, where the signal decomposition algorithm allows
for more than one hit per segment, and the number of
single-interaction points is found to be less than what is expected
from Geant4~\cite{Agostinelli2003250} simulations (see Tab. 6 of ref.
\cite{LAURITSEN201646}). How single-interaction clusters are selected
and validated depends on the algorithm, but the common criterion to
accept or reject the cluster is the depth of the interaction point
inside the AGATA detectors.

\subsection{Limiting factors}
\label{sed:gsfgrtsec:lf}
Physics imposes some limits on \grt:  
\begin{enumerate}
  \item The electron generated by the Compton scattering will deposit
    its energy in a volume of up to several mm$^3$. The energy loss
    proceeds via ionisation and the emission of bremsstrahlung. As a
    consequence, even with an exact PSA, the Compton relation will not
    be perfectly fulfilled as the vertex given by PSA is displaced
    with respect to interaction position.
  \item The relation given in equation (\ref{compt}) is only valid
    when the electrons on which the \gr\ scatters is at rest. This is
    not the case for electrons in atoms as they have finite momentum,
    making the relationship between angle and deposited energy an
    approximation only.
  \item Rayleigh scattering of \grs. As the \gr\ does not loose energy
    but changes direction an uncertainty in the scattering direction
    is introduced. This happens mainly for low-energy \grs, i.e. at
    the end of a track closer to the photo-electric absorption than
    the resolving power of the PSA. In practice its impact is therefor
    small on the performance of \grt.  
  \item For tracks with 2 interactions there is an ambiguity for the
    order of the two interactions. 
\end{enumerate}
These limitations lead to an an overlap between the figure of merit
for correctly ordered fully absorbed \grs\ and not fully absorbed
\grs\ or wrongly ordered sequences of interaction points. This was
recognised early in the AGATA and GRETA projects, e.g. see Milechina
et al.~\cite{MILECHINA2003394} discussing the impact of the finite
electron momentum and Vetter et al. \cite{VETTER2000223} adding to
this the discussion of the finite range of the Compton electron. In
the work of Lopez-Martens et al. \cite{LopezMartens2004454} these
effects were simultaneously quantified using
Geant4~\cite{Agostinelli2003250} simulations. The simulations were
made assuming a position-resolution of about 2.4 mm FWHM for
Milenchina et al. \cite{MILECHINA2003394} and 5 mm FWHM for
Lopez-Martens et al. the \cite{LopezMartens2004454}. Energy
depositions were packed within the assumed position resolution. The
conclusion of Lopez-Martens et al. is that both for back-tracking and
cluster-based tracking algorithms the major uncertainty in the
scattering angle comes from the finite volume in which the Compton
electron deposits its energy. Including the energy-loss process of the
Compton electron reduced the efficiency for fully detecting 30 1.3 MeV
\grs\ using the back-tracking method from 23.4\% to 21.1\%. Including
the electron momentum distribution give 21.9\% and 21.2\%,
respectively. This is compatible with what was found by Milechina et
al.~\cite{MILECHINA2003394}, were the efficiency dropped from 33\% to
24\% for a multiplicity 25 of 1.3 MeV \grs\ when including the
electron momentum. The conclusion is that improving the position
resolution from PSA (presently around 5 mm FWHM
\cite{RECCHIA200960,Soderstrom201196,LJUNGVALL2020163297}) will
improve the \grt.

Hammond et al.~\cite{HAMMOND2005535} pointed out that for \grs\ with
an energy above 255 keV that are absorbed in two interactions, i.e.
one Compton scattering followed by a photo-electric absorption, there
is an ambiguity in the order of the two interaction points. In theory
this is most troublesome for \grs\ with in the energy range of 500-700
keV. In AGATA, that assumes one interaction point per segment, this
ambiguity is found in a larger energy range and introduces an
uncertainty in the ordering of the interaction points.

\section{Gamma-ray tracking algorithms implemented in AGATA}
\label{sec:grta}

The \grt\ algorithms that are used in the AGATA collaboration have
been described in a recent review by Korichi and
Lauritsen~\cite{Korichi2019}. For this reason we will only give an
description of the major changes to the OFT code since then. For other
\grt\ codes and algorithms that are or have been in use with the AGATA
community we only provide short descriptions with the appropriate
references as no significant development has been made.

\subsection{OFT}
\label{sec:grta:sec:oft}

The original OFT code is described in reference
\cite{LopezMartens2004454}. In this section, we will detail the recent
modifications to the code.

\subsubsection{Distance calculation}

\begin{figure}[tbh]
\begin{center}
\includegraphics[width=\columnwidth]{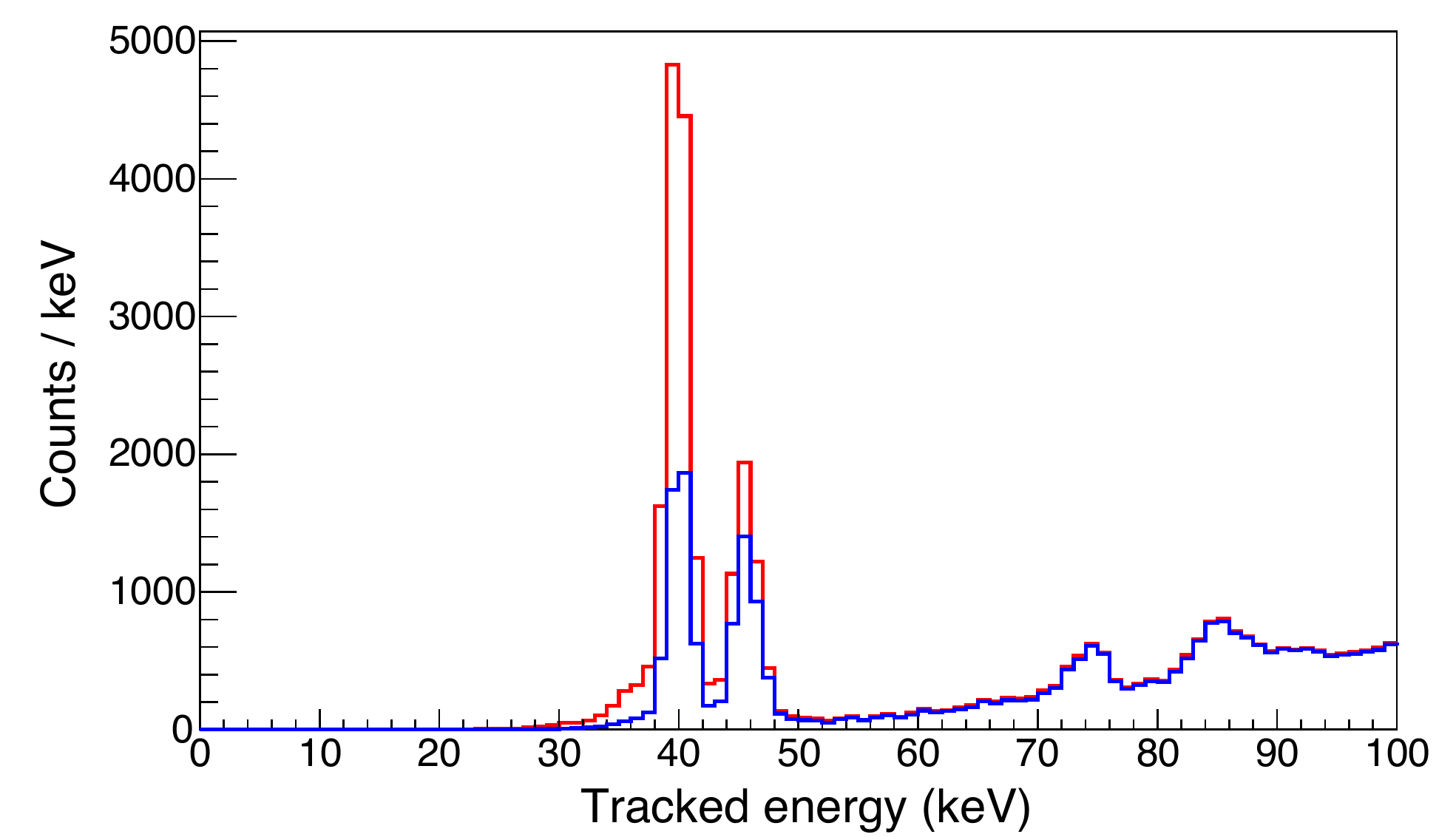}
\end{center}
\caption{Comparison of the low-energy part of a tracked spectrum of
  $\gamma$ rays emitted by a $^{152}$Eu source with the usual sphere
  approximation (blue) and correcting for the overestimation of the
  effective distances travelled in Ge (red). The gain in efficiency at
  40 keV is close to factor 3. The data was taken at GANIL.}
\label{geom_corr}
\end{figure}

To compute the ranges of photons in Ge, effective distances in Ge
between interaction points as well as the effective distance between
every interaction point and the position of the source need to be
calculated. The problem can be solved geometrically if one
approximates the detector geometry to a shell of Ge of inner radius
$R_0$. This approximation was checked with the Geant4 AGATA simulation
code~\cite{Farnea2010331}, in which the exact geometry of AGATA is
defined, i.e. the shape of the crystals, the encapsulations, the
cryostats, the empty spaces and the distances between all these
elements. The efficiency and P/T were found to be the same at 1.3 MeV
using the exact distances as compared to using distances obtained via
the shell approximation. However, the approximation leads to an
overestimation of the distance travelled by photons from the source
into the detector by up to a few mm. The extra distance travelled is
greatest for interactions situated at large radii in the detectors.
This overestimation is extremely penalising for low-energy photons
($<$60 keV), which have very small ranges in Ge and are therefore
awarded a poor range probability by OFT. A correction has been added
to the distance calculation routine in the OFT algorithm. Figure
\ref{geom_corr} shows the resulting improvement of the tracked
efficiency at low energy. For a more in depth discussion of \lq\lq
tracking\rq\rq\ at very low energies, see Section
\ref{sec:pogrtia:lowe}.

\subsubsection{Cluster search}

Points are clusterised according to their relative angular distance.
If the relative angular separation between interaction point $i$ and
any other interaction point is larger than $\alpha$, $i$ is assigned
to a single interaction cluster. For a given value of $\alpha$, no
interaction point can be assigned to more than 1 cluster. This
clusterisation is repeated for various values of $\alpha$. The span of
$\alpha$ depends on the total number of interaction points in the
event $nb\_int$ :
\begin{equation}
\alpha_{max} = acos( 1- \frac{2}{((nb\_int+2)/3)^{0.9}})
\end{equation}
The introduction of a dependence of the maximum allowed value of
$\alpha$ on the number of interaction points in the event was
fine-tuned with simulated data for MGT. For OFT, including such a
dependence in the code resulted in an increase of the simulated
efficiency and peak-to-total for low and medium $\gamma$-ray
multiplicities and a slight improvement of the tracking performance at
high multiplicities. A new feature has recently been added to the
code, namely the possibility to reduce $\alpha_{max}$ by a factor to
be fine-tuned by the user. As an example, reducing $\alpha_{max}$
increases the efficiency to track 100-300 keV $\gamma$ rays for events
with a small number of interaction points.

\subsubsection{Cluster evaluation}

\underline{Compton events}\\

In OFT, the figure of merit $L$ of a particular sequence of $n$
interaction points is given by:
\begin{equation}
\label{FOM}
L^{2n-1} = \prod_{i=1}^{n-1} P_i \exp ^{-a \left(\frac{V_i^E}{\sigma_E}\right)^2}
\end{equation}
where $i $ is the Compton vertex number, $P_i$ contains the physics
information regarding interaction probabilities at $i$ and $i+1$ and
ranges in Ge to and from $i$, $a$ is 2 at $i=1$ and 1 for $i>1$ and
$\sigma_E$ is the uncertainty in the determination of the scattered
energies due to the uncertainty in the determination of
interaction-point energy depositions as well as positions. The average
position uncertainty is parameterized by a free parameter
$\sigma_{\theta}$ in cm and it enters in the calculation of
$\sigma_{E}$. Until the availability of data from AGATA, OFT was
developed with simulated data sets produced with the AGATA simulation
code (see Farnea at el. \cite{Farnea2010331} and Labiche et al.
\cite{Labiche2023}in this issue). The output of the simulations was
modified to mock the expected experimental conditions, such as energy
resolution and threshold and position resolution. To reproduce the
experimental position resolution, interaction points were packed
together if they happened to be in the same segment of the same
detector and their positions smeared in $x,y$ and $z$ using a Gaussian
distribution of full width at half maximum (FWHM) :
\begin{equation}
FWHM(cm) = S_0\times\sqrt{\frac{0.1}{e_i}} 
\label{FWHM}
\end{equation}

where e$_i$ is the energy of interaction point $i$ in MeV and $S_0$
was taken to be 0.5 cm. The optimal parameters of the algorithm were
tuned in such a way as to maximise the product of efficiency and P/T
and the best tracking performance was obtained with
$\sigma_{\theta}$=0.24 cm in the case of 1 MeV incident photons
\cite{LopezMartens2004454}. The experimentally-determined value of
$S_0$~\cite{Soderstrom201196} turns out to be larger and this explains
why with real source and in-beam data, the optimal parameter
$\sigma_{\theta}$ was extracted to be 0.8 cm. It was also found that
setting $a$=2 for the first Compton vertex in the track reduces the
performance of OFT compared to what was obtained with simulated data
and so now $a$ is set to 1 for all vertices.\\

\underline{Pair-production events}\\

To increase the tracking efficiency at high energy, it is necessary to
track pair-production events since for photon energies above 10 MeV,
pair-production is the dominant physical interaction process in Ge.
However, it must be considered that including pair production becomes
important already at \gdr\ energies significantly lower than 10 MeV,
since we are interested only in full energy peak events. The pattern
that is tracked is the one in which a cluster contains an interaction
point collecting the incident photon energy $E_{\gamma}$ minus twice
the electron rest-mass.

\begin{figure}[tbh]
\begin{center}
\includegraphics[width=\columnwidth]{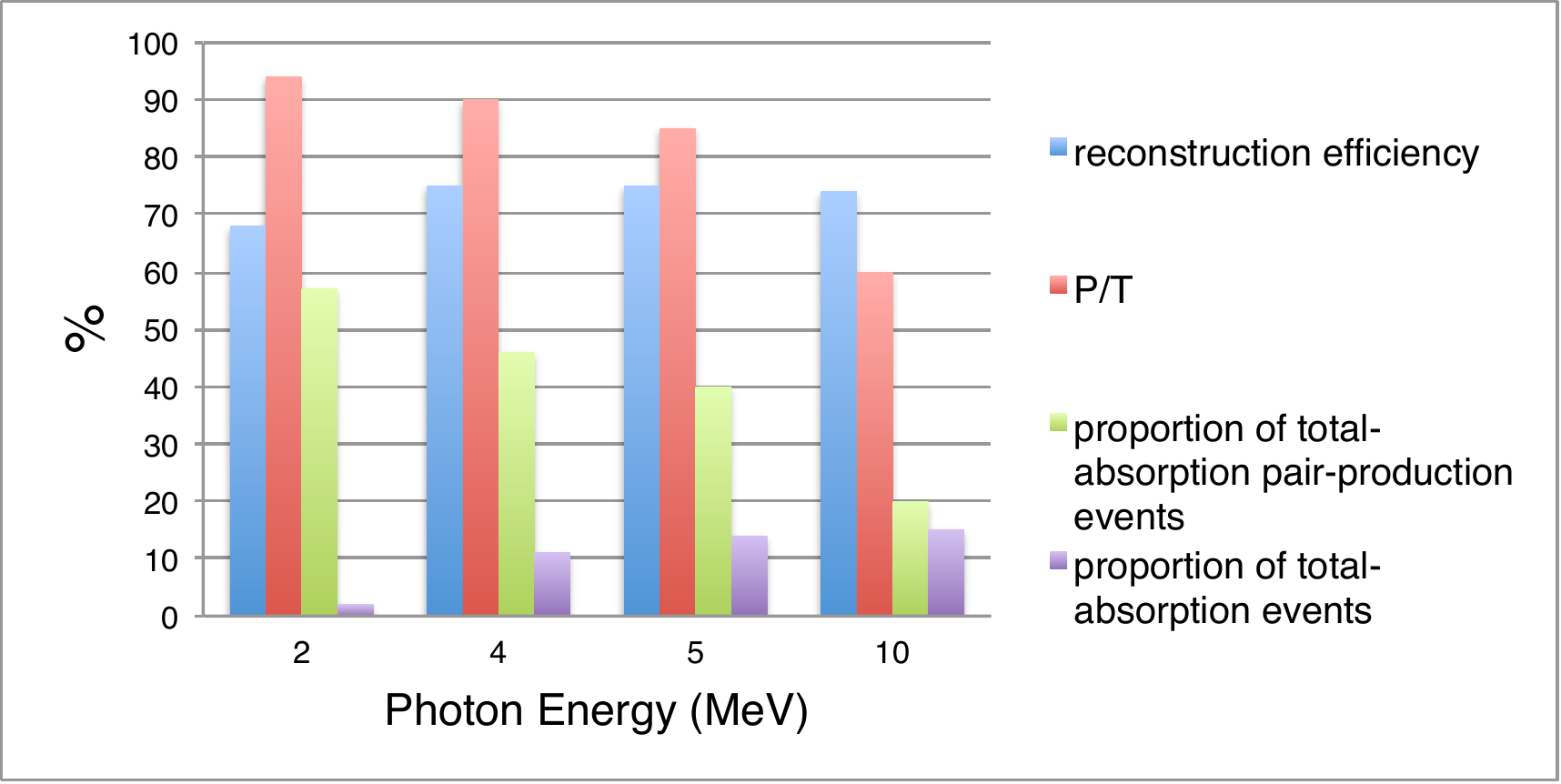}
\end{center}
\caption{Reconstruction efficiency and P/T of the pair-production
  tracking routine as a function of incident photon energy. The
  fractions of the pair-production and overall events resulting in
  total absorption, which can, in principle, be tracked are also
  shown. Data from geant4 simulations of HPGe shell and treated as in
  Lopez-Martens et al. \cite{LopezMartens2004454}.}
\label{statpair}
\end{figure}

The total-absorption events of this type represent $\sim$60$\%$ of the
total-absorption pair-production events for 2 MeV incident photons
(see Fig.~\ref{statpair}). This number decreases steadily as the
incident photon energy increases to reach roughly 20$\%$ at 10 MeV.
This decrease is due to the increase in energy transferred to the
electron-positron pair, which results in a less localised deposition
of energy around the position of the pair-production interaction
point. However, this type of event represents a growing fraction of
the overall total absorption events (2$\%$ at 2 MeV, 10$\%$ at 4 MeV
and close to 15$\%$ at 10 MeV). This trend is related to the increase
of the pair-production cross-section with increasing incident photon
energy.
To accommodate for pair-production clusters, the maximum number of
allowed interaction points in the clusterisation procedure has been
increased from 7 to 10, but clusters with more than 7 interaction
points are not evaluated as possible Compton events since this
deteriorates the global P/T.

If a cluster has an interaction point with an energy corresponding to
the total cluster energy minus twice the electron rest-mass it is
given the figure of merit equal to the square root of the probability
for the incident photon to travel the distance in Ge from the source
to the position of the interaction point multiplied by the probability
to undergo a pair-production interaction. If this figure of merit is
found to be larger than the best Compton sequence, the cluster is
flagged as a pair-production event.

As can be seen in Fig.~\ref{statpair}, the reconstruction efficiency
of such pair-production events is rather high, since it is of the
order of 70$\%$ for incident photons of energy ranging from 2 to 10
MeV. Furthermore, the number of wrongly recognised pair-production
clusters is small: the peak-to-total of the pair-production tracking
routine is 85$\%$ at 5 MeV, and 60$\%$ at 10 MeV. Reconstruction of at
least one of the 511 keV annihilation photons did not improve the P/T
and led to an overall reduced pair-production tracking efficiency.\\

\underline{Single-interaction events}\\

The evaluation of single-interaction-point clusters is performed after
the evaluation of the multi-interaction-point clusters.

One of the criteria for a \lq\lq clean\rq\rq\ identification of a
single-interaction point is that it is well isolated from other hits.
The optimal minimal distance to the closest interaction point is found
to be 4 cm. If a single-interaction point fits this criterion, the
figure of merit of the corresponding cluster is computed. In reference
\cite{LopezMartens2004454}, it was defined as the square root of the
probability for the incident photon to travel the distance in Ge from
the source to the position of the interaction point multiplied by the
probability to undergo a photoelectric interaction. Depending on the
threshold of acceptance, the spectrum of \lq\lq
tracked\rq\rq\ single-interaction clusters had a different end point:
$\sim$ 600 keV for a minimum figure of merit of 0.15 and $\sim$ 2 MeV
for 0.02. Accepting high-energy single interaction points therefore
came at the cost of a large background. The procedure has now been
changed and only the range in Ge is considered through an
experimentally-defined energy-dependent formula (a second order
polynomial expression). This results in a reduction of the background
at low energies while preserving the photo peak efficiency.

\subsection{Other \grt\ algorithms}
\label{sec:grta:sec:other}

\begin{itemize}
\item The MGT code~\cite{BAZZACCO2004248,bazzaccoMGT,Kroll2003} was
  developed in the TMR program \lq\lq Development of $\gamma$-ray
  tracking detectors\rq\rq. It was used extensively in the early phase
  of AGATA.
\item The tracking code used by the GRETINA/GRETA~\cite{Schmid199969}
  collaboration, referred to as GT, has also been tested with AGATA
  data. The two tracking codes, i.e. OFT and GT, perform very similar
  with a slight advantage at high \gdr\ multiplicities for OFT and at
  low \gdr\ multiplicities for
  GT~\cite{Korichi2019,Korichiinprep,lopezagatagretina}.
\item Although not a part of the official distribution of software of
  AGATA, the Bayes tracking~\cite{Napiralla2019EmployingT} has been
  tested on source data, and is included as an algorithm that has been
  implemented for AGATA. The algorithm formulates the problem of
  \gdr\ tracking, with the help of Bayes theorem, as maximising the
  probability for the measured set of interaction points given a
  number of emitted \grs. This for all possible permutation, for 1 to
  N \grs\ where N is the number interaction points. For such a
  rigorous mathematical formulation estimation of probabilities for,
  e.g., the number of interaction for a \gr\ of given energy, have to
  be calculated. This has been done using
  Geant4~\cite{Agostinelli2003250} simulations with the standard AGATA
  code\cite{Farnea2010331}, i.e. using packed simulated data with
  realistic detector geometry. The Bayes tracking algorithm can in
  principle also reconstruct the \gdr\ energy for a not fully absorbed
  \gr.
\end{itemize}

\section{Other clustering and $\gamma$-ray tracking algorithms}
\label{sec:ogrta}
Extensive investigations have been made to try to improve the
clustering+validation process used in the MGT, OFT, and GT algorithms.
These efforts can be classified in two groups whether it is the
clustering or the evaluation of the clusters that is targeted. The
Fuzzy C Logic~\cite{Suliman2010} and the Deterministic Annealing
Filter~\cite{DIDIERJEAN2010188} aim at improving the clustering of
\gr\ interaction points into trail clusters before evaluation. Both
have been tried with simulated data giving interesting results. The New
tracking ALGOrithm for gamma rays (TANGO)~\cite{TASHENOV2010592}
introduces more complex figure-of-merits in the cluster evaluation and
the possibility to estimate the energy of not fully absorbed \grs.

A first approach to a self-calibrating $\gamma$-ray tracking
algorithm, based on experimental data, was presented
in~\cite{NAPIRALLA2020163337}. In this paper the influence of
non-Gaussian behaviour in scattering angles between three points in
3D-space as well as the impact of PSA-induced interaction point
merging on Compton-scattering angles and consecutively on $\gamma$-ray
tracking with AGATA have been discussed. Although the algorithm was
tested with $^{137}$Cs source data, for its application in real
experimental conditions and for having a thorough comparison with the
performances of OFT tracking, the study still needs completion.
Gamma-ray tracking was used in the work by Heil et al.~\cite{Heil2018}
to experimentally determine a pulse-shape data base for PSA.
Pulse-shapes from experimental data are grouped together into
collections of hits, based on their pulse shapes. Each hit collection
is then given a position. All the hits inside a hit collection are then
linked to the hits in the other hit collections. The positions for the
hit collections are then varied to maximise the Figure-Of-Merit,
corresponding to \grt\ of the linked hits. As each hit collection
contains many different hits linked to other hits the location of the
hit collections have well optimal positions. The optimisation
procedure is performed in an iterative manner. It has to date only
been tested on simulated data using the actual positions of the
interaction as the criteria to create the hit collections. Efforts to
test the method with experimental data and pulse-shapes are ongoing
(end of year 2022).

Andersson and B{\"a}ck \cite{ANDERSSON2023168000} have recently
investigated the use of graph neural networks for \grt. This
exploratory work based on Geant4 simulations produced very encouraging
results where the neural network based tracking outperforms both
forward tracking and back tracking on simulated data. In the
simulations an ideal $4\pi$ Ge shell was used.

\section{Performance of $\gamma$-ray tracking in AGATA}

The performance of AGATA has been continuously evaluated since its
first materialisation as the AGATA demonstrator at Legnaro-INFN
Laboratory \cite{Gadea201188}. These evaluations have focused on
Photopeak efficiency ($\epsilon_{ph}$) and the Peak-to-Total (P/T) of
both the crystals used as individual detectors and after \grt. By
looking at both the individual detectors and the final result of \grt,
and comparing them with Geant4 simulations, it is possible to not only
characterise the performance of AGATA but also to find where
improvements might be possible. In the following these findings will
be discussed individually for three different energy regimes. The
\lq\lq normal energy regime\rq\rq\ will be discussed in section
\ref{sec:pogrtia:normale}, the \lq\lq low energy regime\rq\rq\ is
discussed in section~\ref{sec:pogrtia:lowe}, and finally the \lq\lq
high-energy regime\rq\rq\ in section~\ref{sec:pogrtia:highe}.

\subsection{Normal $\gamma$-ray energy regime
  ($\approx 100~keV<E_{\gamma}<5~MeV)$}
\label{sec:pogrtia:normale}

This energy regime is characterised by the dominance of Compton
scattering. The pair production mechanism becomes of relevance toward
the upper energy boundary of the interval. Tracking of events with
energy close to the lower energy boundary is limited by the mean-free
path of the $\gamma$ rays being only a few millimeters and the low
signal-to-noise ratio of the pulses not allowing the PSA to identify
the interaction positions with high precision.

The performance of AGATA at GSI~\cite{Lalovic2016258} as well as at
GANIL~\cite{LJUNGVALL2020163297} in this energy regime were thoroughly
investigated using source measurements. Resulting efficiencies and
P/T's were compared with simulations. For the GSI measurements 21
crystals were present, whereas for the GANIL measurements 30 detectors
were installed in the array, out which one were not used for the
efficiency measurements. In both works the absolute efficiency after
\grt\ were determined using standard \gdr\ sources. \lalovic\ et al.
used \nuc{60}{Co}, \nuc{152}{Eu}, and \nuc{56}{Co} whereas Ljungvall
et al. did not look at the high-energy response given by \nuc{56}{Co}.
Efficiencies were determined using both singles measurements
corrected for dead time as well as coincidence measurements corrected
for angular correlations for both the GSI and GANIL setups.  

\begin{figure}[tbh]
  \begin{center}
    \begin{subfigure}[t]{.99\columnwidth}
      \centering
      \includegraphics[width=.99\columnwidth,trim=10.5cm 13.3cm 1.5cm 7cm,
        clip]
                      {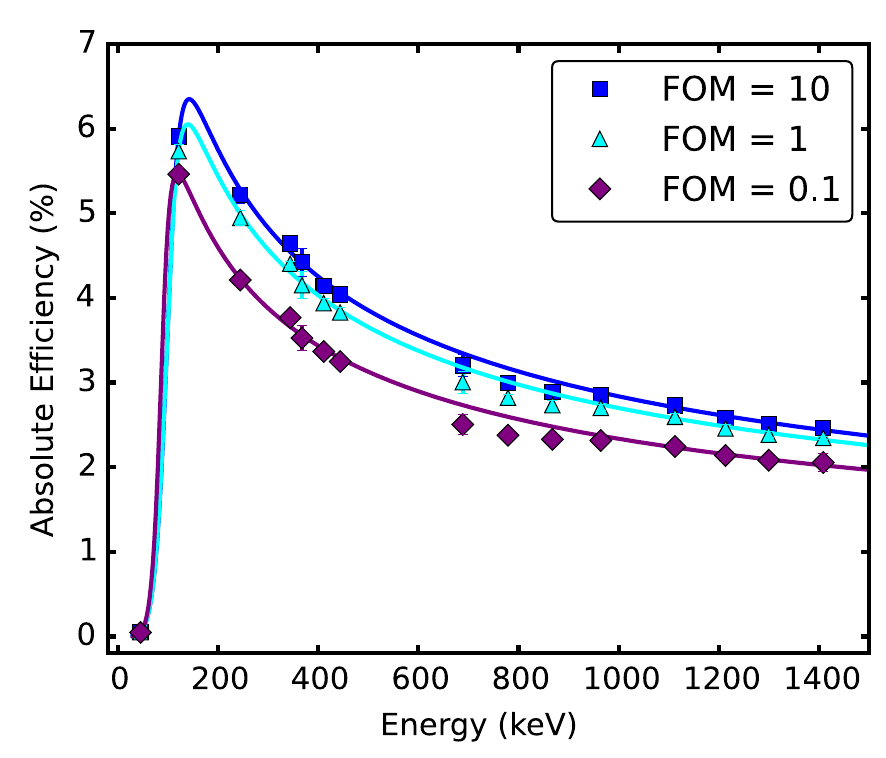}
                      \caption{Efficiency after \grt\ as a function
                        \gdr\ energy for different the Figure-Of-Merit
                        (FOM) used in MGT to decide if a cluster of
                        interaction points is correctly tracked. A
                        FOM=1 is the default used in MGT. 
                        Figure taken from \lalovic\ et al.
                       ~\cite{Lalovic2016258}}
                      \label{effagatagsi}
    \end{subfigure}
    \begin{subfigure}[t]{0.99\columnwidth}
      \centering
      \includegraphics[width=.99\columnwidth,trim=0cm 0cm 0cm 0cm,
        clip]
                      {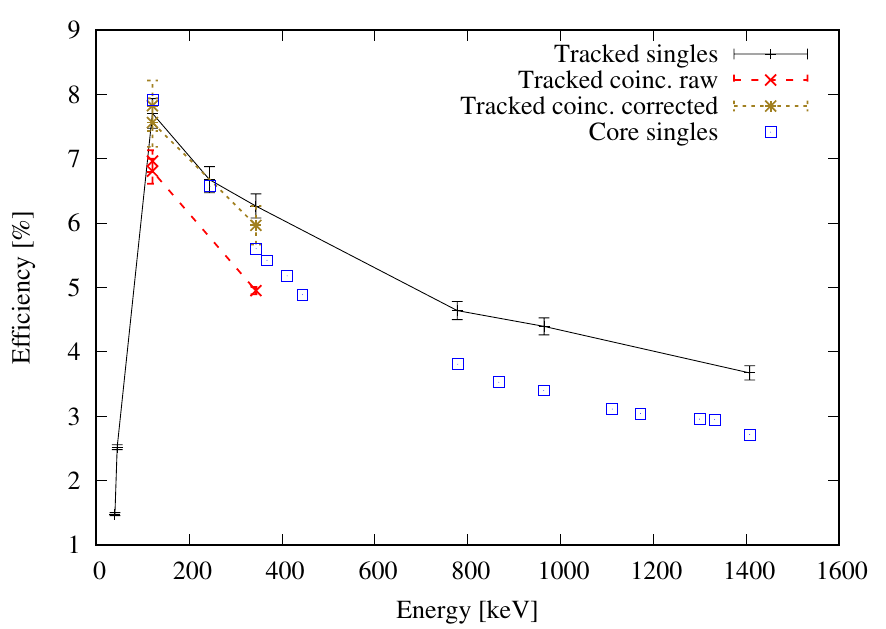}
                      \caption{Efficiency after \grt\ as a function
                        \gdr\ energy for singles and coincidence
                        measurements using the OFT code. The
                        efficiency using the sum of crystal (called
                        core singles) is also shown. Figure from
                        Ljungvall et al.~\cite{LJUNGVALL2020163297}.}
                      \label{effagataganil}
    \end{subfigure}
  \end{center}
  \caption{Efficiency after \grt\ for AGATA at GSI (a) and AGATA at
    GANIL(b), respectively. For GSI the MGT~\cite{BAZZACCO2004248}
    code was used for tracking. The OFT~\cite{LopezMartens2004454} code
    was used for the GANIL data.}
\end{figure}

The measured efficiencies after \grt\ at 1.4 MeV are 2.50(2)\% and
3.67(1)\% for the GSI and GANIL phase, respectively. The P/T is the
two cases are 38(1)\% and 36(1)\% respectively. A linear scaling from
21 to 29 detectors would give an efficiency of 3.5\%. However, these
efficiency numbers depend on the choice of parameters both for MGT
and OFT so the two set of data taken at different times are indeed
compatible with each other. The efficiencies as a function of
\gdr\ energy are shown in figure~\ref{effagatagsi} for GSI and in
figure~\ref{effagataganil} for GANIL, respectively. In all cases
single-interaction events are included.

\subsection{Low-energy $\gamma$ rays ($E_{\gamma}<100 keV$)}
\label{sec:pogrtia:lowe}

For \grs\ with an energy lower than about 100 keV \grt\ is not really
applicable as the probability for scattering is very low. Instead the
\lq\lq single interaction validation\rq\rq\ procedure is used. As
described in section~\ref{sec:grta:sec:oft}, this is essentially a
test comparing the distance the \gr\ has travelled in germanium with
an empirical relation deduced from experimental data. For low-energy
\grs\ it is also important to correctly calculate the path inside the
HPGe crystal, as shown in section~\ref{sec:grta:sec:oft} and
figure~\ref{geom_corr}.

In the case of very-low energy \grs\ and x-rays an additional
complication arrives. The very low signal-to-noise ratio give
unreliable PSA - the interaction is therefore often positioned by the
PSA much deeper into the crystal than it occurred (this can be
evidenced by looking at the depth distribution of interaction points
as given by PSA gating on an, e.g., x-ray line in $^{152}$Eu using
non-tracked data). The single interaction validation procedure will
then discard it as a low-angle Compton scattering of a higher energy
\gr. In the case that low-energy efficiency is needed a solution has
been proposed in which under certain conditions the PSA result is
\lq\lq corrected\rq\rq\ to a value that is compatible with the
mean-free path of a \gr\ with the energy of the interaction. For
further details see Cl{\'e}ment et al.\cite{Clement2022}.

\subsection{High-energy $\gamma$ rays ($E_{\gamma}>5 MeV$)}
\label{sec:pogrtia:highe}

\begin{figure}[htp]
\begin{center}
  \includegraphics[width=0.88\columnwidth]{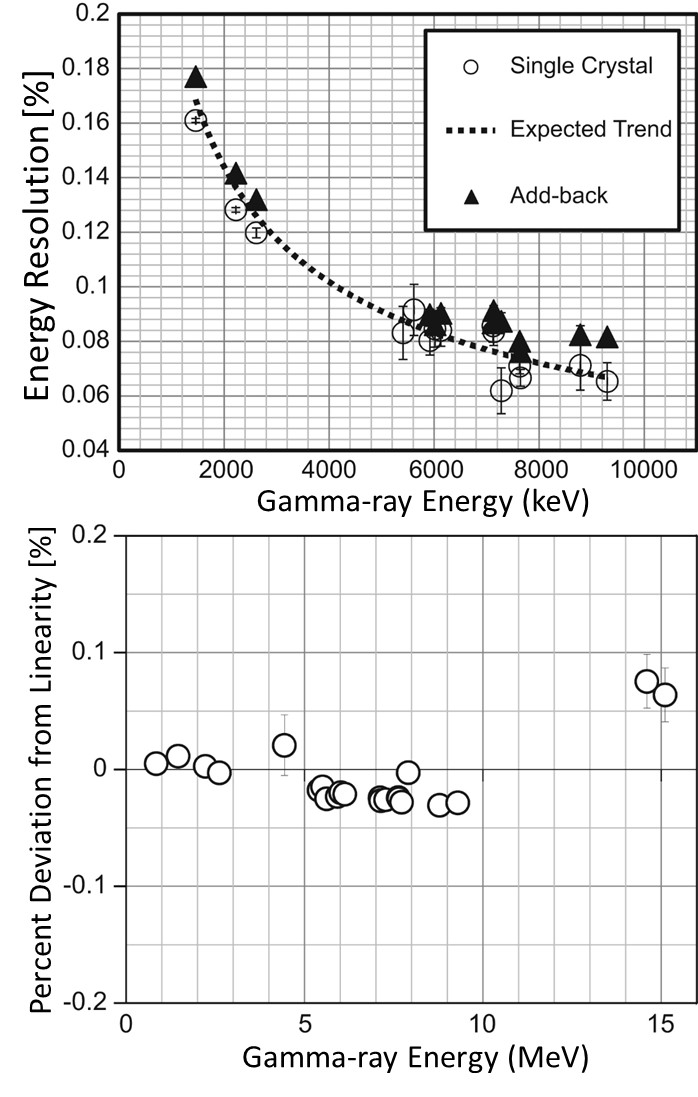}
  \end{center}
\caption{Upper panel: energy resolution of the AGATA detectors up to 9
  MeV. The data for the best performing HPGe crystal are shown by
  empty black circles. The black triangles represent the energy
  resolution for the add-back procedure (sum of the energies recorded
  in all the crystals that fired in each event). The experimental data
  follow the expected $\propto$ $E^{-1/2}$ trend (indicated by the
  dashed black line). Bottom panel: percent deviation of the measured
  energies from the tabulated ones, considering a linear calibration.
  If not displayed, error bars are smaller than the symbol size.}
\label{pscstheo1}       
\end{figure}

It is worth mentioning that an idea to improve the performances at
high energies, for $\gamma$-ray spectrometers based on position
sensitive detectors, was already proposed rather long time ago by
Glenn Knoll and collaborators~\cite{1221927}. In the case of AGATA, a
work of 2013~\cite{Crespi2013} studied the response of AGATA
detectors to $\gamma$ rays up to 15.1 MeV. Gamma rays up to an energy
of $\approx$9 MeV were obtained with an extended Am-Be-Fe source.
Then, the only feasible way to have $\gamma$ lines of even higher
energies was to use in-beam reactions (see e.g.
\cite{MILLION2000422,CIEMALA200976}). In particular, the reaction
d(\nuc{11}{B},n$\gamma$)\nuc{12}{C} at E$_{\text{beam}}$=19.1 MeV was
used to produce the 15.1 MeV $\gamma$'s. To be noted that in this case
the $\gamma$'s are emitted in flight by the \nuc{12}{C} ions
($\beta\approx4\%$) and this fact was used to test the Doppler
correction capabilities of AGATA at these high energies. The energy
resolution of AGATA detectors was found to scale $\propto$ $E^{-1/2}$
up to 9 MeV, as expected (see upper panel of Fig.~\ref{pscstheo1}).
Also the linearity was studied and the energy-to-pulse-height
conversion resulted to be linear within $\sim$0.05\% up to the
$\gamma$ energy of 15.1 MeV (see lower panel of Fig.~\ref{pscstheo1}).

These data also showed that the application of $\gamma$-ray tracking
allows some suppression of background caused by n-capture in Ge
nuclei. The neutrons, in this case, were emitted by the Am-Be source.
This background suppression capability of the tracking software can be
directly appreciated observing the disappearance of the 10.196 MeV
peak originated indeed from neutron interactions (see Fig. 8
in~\cite{Crespi2013}). The value 10.196 MeV corresponds, in fact, to
the sum-energy of the $\gamma$'s emitted following the \nuc{74}{Ge}
nucleus de-excitation (Q-value of the neutron capture reaction).
\\ From the operational point of view, it was important to properly
set the AGATA electronics for the detection of high-energy $\gamma$
rays. Specifically, in order to have the $\sim$20 MeV dynamic range
also for the segment signals acquisition (normally the $\sim$4 MeV
range is set). The core, instead, has always two channels in the data
acquisition by default, for the $\sim$4 MeV and $\sim$20 MeV range
respectively. \\ The Doppler correction quality, at the very high
energy of 15.1 MeV, was found to be consistent with the expectations,
according to a dedicated Geant4 simulation (see results reported in
in~\cite{Crespi2013}). The main limiting factor in the Doppler
correction quality, in this case, was found to be the missed
event-by-event reconstruction of the velocity vector of the
\nuc{12}{C} nucleus. From an extrapolation based on the results
presented in~\cite{Crespi2013}, an intrinsic resolution of about 10
keV is expected at the $\gamma$ energy of 15 MeV.\\ Regarding the
tracking performances (MGT algorithm was used) at 15 MeV, it was found
more convenient to use the so-called calorimeter mode, for getting the
non-Doppler corrected energy of an event, and then use the position of
the most energetic hit given by the PSA algorithm, for determining the
incoming direction of the $\gamma$ ray, to be used in the Doppler
correction formula (this method was named "PSA+1HitID"). It has to be
specified that this in-beam test was performed at LNL in 2010: only
two clusters of AGATA were present during the experiment and
significant improvement in the treatment of pair production events in
the tracking software was made after that time. Moreover, the fact
that the "PSA+1HitID" works better than the standard tracking is
strictly connected to the $\gamma$ multiplicity (M=1) situation and
the minimal presence of background radiation. In these specific
conditions, the presented results might suggest a simple and efficient
alternative to standard tracking.

\subsection{Performance of $\gamma$-ray tracking with AGATA coupled
  to different detector systems }
\label{sec:pogrtwactdds}

To increase the sensitivity of the experimental setup
\gdr\ spectrometers are often coupled to other detector systems.
Examples are the use of magnetic spectrometers
\cite{STEFANINI2002217,Rejmund2011184} for event-by-event
identification of the reaction product. Particle detectors to count
charged particles and/or neutrons originating from the reaction are
also used to enhance the wanted reaction channel in the \gdr\ spectra
\cite{SCHEURER1997501,VALIENTEDOBON201981,ASSIE2021165743}. The use
these ancillary detectors with AGATA is described in previous papers
\cite{Gadea201188,DOMINGOPARDO2012297,CLEMENT20171} and will not be
further discussed here, and we will focus on how different categories
of ancillary detectors impact the \grt\ performance.

\begin{figure}[tbh]
  \begin{center}
    \includegraphics[width=.99\columnwidth,trim=0cm 0cm 0cm 0cm,
      clip]{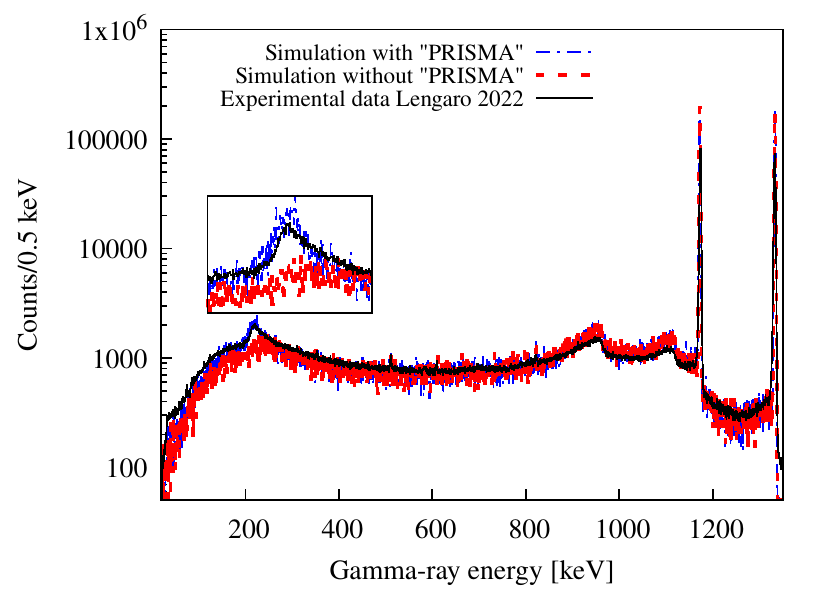}
      \caption{Tracked spectra using a \nuc{60}{Co} source. The solid
        black line is the experimental spectrum, the dashed red line
        is a simulation without an iron block, and the dashed dotted
        blue line is the simulation with an iron block modelling
        PRISMA. The appearance of the back-scattering peak in the
        simulated data with PRISMA is clear, as shown in the inset.
        For details see text.}
      \label{fig:bumpfromspectros}
  \end{center}
\end{figure}

Large acceptance magnetic spectrometers have an impact on the spectra
after \grt. As shown in figure~\ref{fig:bumpfromspectros} a large
back-scattering peak is present in the experimental data that is not
present in simulated data unless a large block of steel in introduced
in the simulation. This block of steel represents the entrance
quadrupole magnet of PRISMA~\cite{STEFANINI2002217}. All
\grt\ algorithms have problem discriminating against these back
scattered \grs\ as they come from a direction close to that of the
target position. The back scattered \grs\ are more present in
\grt\ arrays than in arrays using Compton Shields because the
\grt\ offers less effective collimation than the Compton Shields.

\begin{figure}[tbh]
  \begin{center}
      \includegraphics[width=.99\columnwidth,trim=-.30cm 0cm 0cm 0cm,
        clip]{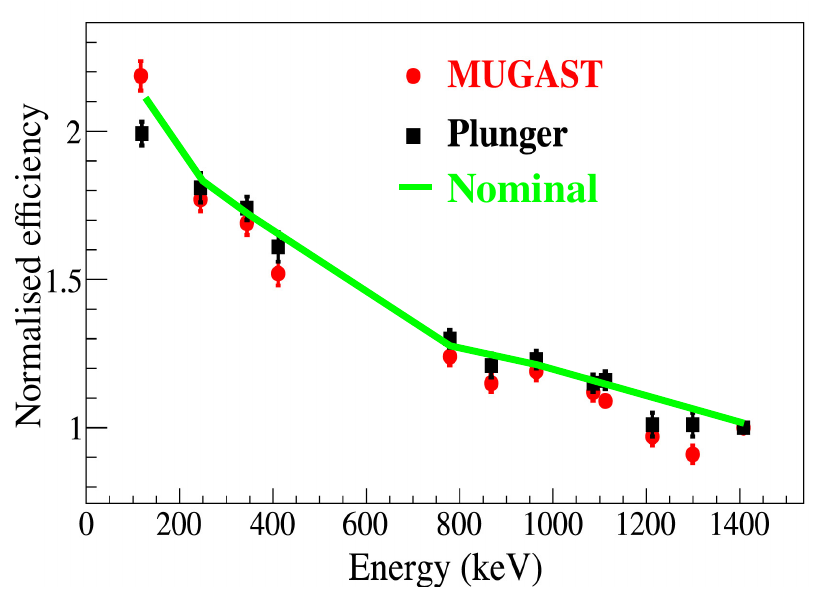}
      \caption{Efficiency, normalised to 1 at 1408 keV, as a function
        of \gdr\ energy. For experimental setups with the OUPS and
        corresponding chamber, the MUGAST particle detector array, and
        finally, with only the standard GANIL target chamber. Figure
        modified from Assi{\'e} et al.~\cite{ASSIE2021165743} and
        Ljungvall et al. ~\cite{LJUNGVALL2020163297}. For errors on
        the nominal efficiency see figure~\ref{effagataganil}.}
      \label{fig:effWithMugast}
  \end{center}
\end{figure}

The use of ancillary particle detectors or other types of devices such
as a Plunger that are positioned between the target and AGATA will
generate scattering and absorption of the emitted \grs. However, for
situations where AGATA only covers a fraction of the solid angle it is
possible to minimise these losses using reasonable designs. Examples
of losses in efficiency is given in figure~\ref{fig:effWithMugast} for
the case of using the OUPS plunger or the MUGAST\cite{ASSIE2021165743}
detector system. The effect in these cases, with AGATA covering less
than $1\pi$ of solid angle, is limited to low energies and moderate in
magnitude.

\FloatBarrier

\section{Improvements in $\gamma$-ray spectrometer performance due to the
  $\gamma$-ray tracking}
\label{sec:ingrspthgrt}

In this section we focus on AGATA performances in respect to a few
specific aspects, in which the detector millimetric position
sensitivity and the tracking technology are of fundamental importance
in improving the experimental sensitivity of a high-resolution $\gamma$
spectrometer.

\subsection{Doppler correction}
\label{sec:ingrspthgrt:sec:dc}

The primary feature of HPGe $\gamma$ detectors is their excellent
energy resolution. For in-beam spectroscopy experiments it can be
significantly degraded due to the Doppler effect. This is especially
important for experiments done in inverse kinematics, which is often
the case for radioactive beam experiments. As Doppler correction is
one of the most important gains that PSA and \grt\ offers, we remind
the reader that, when the $\gamma$ rays are emitted in-flight by a
recoiling nucleus, the width of peaks in the Doppler-corrected spectra
will depend on three factors, namely the intrinsic detector energy
resolution, the error on the velocity vector of the emitting nucleus
and the uncertainty on the photon direction. The last factor depends
on the position resolution of the PSA algorithm used and the capacity
of the \grt\ algorithm to correctly determine the first interaction
point.

The Doppler-shift formula is the following:
  \begin{equation} \label{eq:Doppler}
  E_\gamma^{cm} = E_\gamma
  \frac{1-\beta\cos{\theta}}{\sqrt{1-\beta^2}}
  \end{equation}
where $E_\gamma^{cm}$ is the CMS energy of the $\gamma$ ray,
$E_\gamma$ is the energy of the photon in the laboratory (in other
words the energy seen by the detector), $\beta$ is the velocity of the
emitting nucleus and $\theta$ is the angle between the direction of
the recoiling nucleus and the direction of the photon in the
laboratory. Each of the parameters entering the formula contributes to
the final uncertainty. Quantitatively, the contribution of each
parameter to the final position resolution is evaluated through the
propagation of errors on $E_\gamma^{cm}$, giving:
  \begin{eqnarray} \label{eq:Dopplerbroad}
  \left(\Delta E_\gamma^{cm}\right)^2  & = & {}
    \left(\frac{\partial E_\gamma^{cm}}{\partial\theta}\right)^2 (\Delta \theta)^2 + {}\nonumber \\*
    & + & {}\left(\frac{\partial E_\gamma^{cm}}{\partial\beta}\right)^2 (\Delta \beta)^2 +  {}\nonumber \\*
    & + & {}\left(\frac{\partial E_\gamma^{cm}}{\partial E_\gamma}\right)^2 (\Delta E_\gamma)^2.
  \end{eqnarray}
In this calculation, the different broadening sources are considered
as statistically independent contributions, neglecting for simplicity
any correlation between them. In Eq. \ref{eq:Dopplerbroad}, $\Delta
\beta$ and $\Delta \theta$ are respectively the uncertainty on the
velocity module and on the angle between the direction of the nucleus
emitting the radiation and the emitted \gr. Even if the recoil
velocity vector can be measured on an event-by-event basis, $\Delta
\beta$ and $\Delta \theta$ will be generally non-zero. The term
$\Delta E_\gamma$ in Eq. \ref{eq:Dopplerbroad} describes the
contribution of the intrinsic energy resolution of the detector.

\begin{figure}[htb]
  \centering
  \includegraphics[width=\columnwidth]{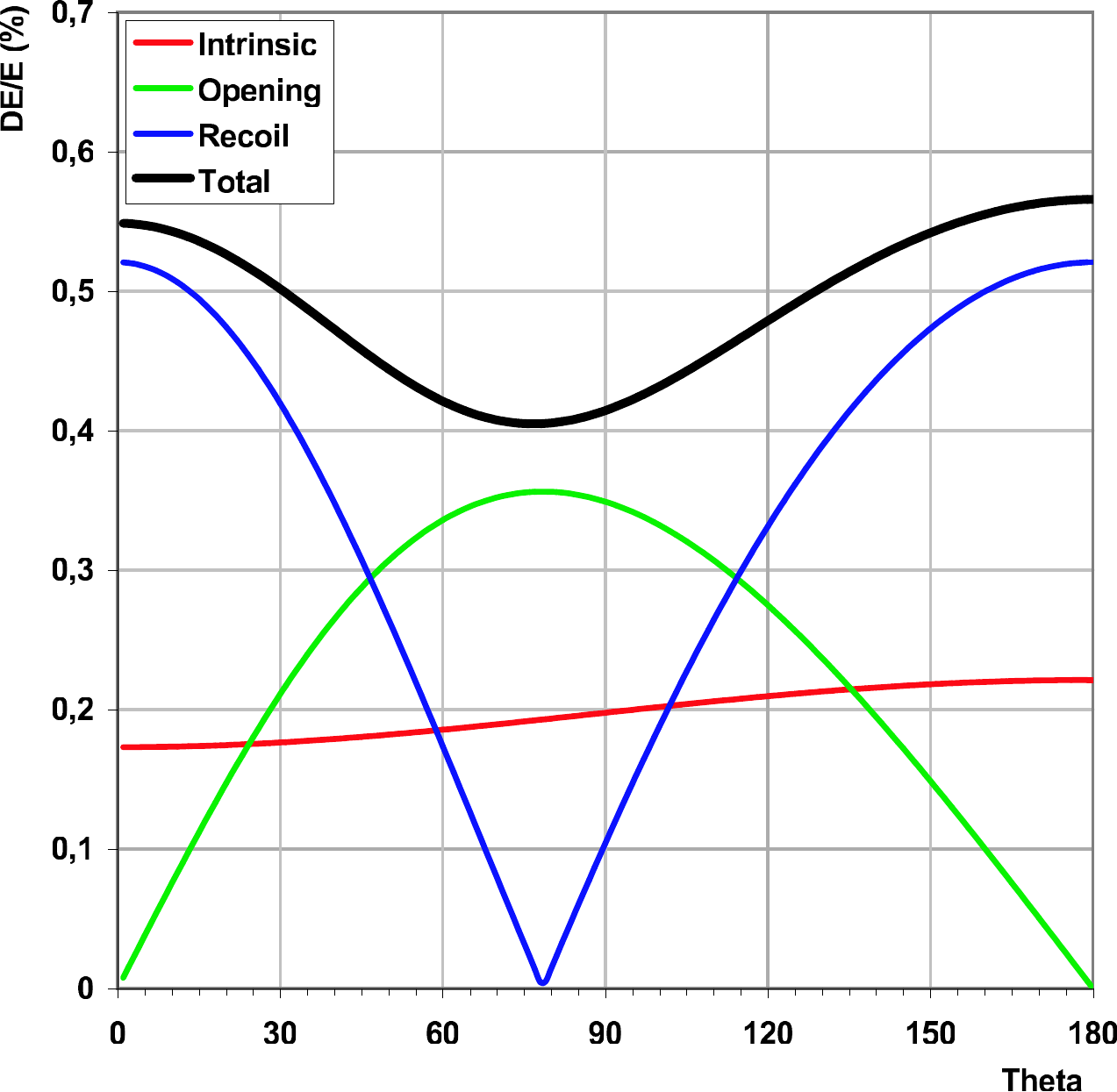}
  \caption[Effects on broadening of \gr\ peak in in-beam
    \gr\ spectroscopy]{ The contributions of the different Doppler
    broadening sources as a function of the azimuthal angle of the
    detector with respect to the direction of the recoil emitting the
    radiation. A photon energy of 1~MeV is assumed, with a typical
    energy resolution for a germanium detector, producing the \lq\lq
    Intrinsic\rq\rq\ contribution (in red); a source velocity of
    $\beta=20.0\%$ with an error of 0.5\%, giving the \lq\lq
    Recoil\rq\rq\ contribution (in blue); an uncertainty $\Delta
    \theta = 1^\circ$ in the source direction, obtaining the \lq\lq
    Opening\rq\rq\ contribution (in green). Taken from
    \cite{Recchia2022}. }
  \label{fig:Doppler}
\end{figure}

The partial derivatives are:
  {
  \begin{eqnarray} \label{eq:DopplerbroadD}
    \frac{\partial E_\gamma^{cm}}{\partial\theta}    & = & {} E_\gamma\frac{\beta\sin{\theta}}{\sqrt{1-\beta^2}} {}\nonumber \\*
    \frac{\partial E_\gamma^{cm}}{\partial\beta}     & = & {} E_\gamma\frac{\beta - \cos{\theta}}{\left(1-\beta^2\right)^{3/2}} {}\nonumber \\*
    \frac{\partial E_\gamma^{cm}}{\partial E_\gamma} & = & {} \frac{1- \beta \cos{\theta}}{\sqrt{1-\beta^2}}
  \end{eqnarray}
  }
The angular error is propagated to the error in the determination of
the CMS energy of the $\gamma$ ray by the coefficient given in the
first raw of Eq.~\ref{eq:DopplerbroadD}. As an example, the
contributions of the three sources of Doppler broadening are sketched
in Fig.~\ref{fig:Doppler}, for the case of photons of 1\,MeV emitted
from a nucleus in motion with $\beta=20\%$ and detected with an
uncertainty $\Delta \theta = 1^\circ$ on its direction. It is thanks
to PSA and \grt\ that a $\Delta \theta $ as low as $1^\circ$ is
imaginable while still keeping such large \gdr\ efficiency.

\begin{figure}[htp]
\begin{center}
  \includegraphics[width=\columnwidth]{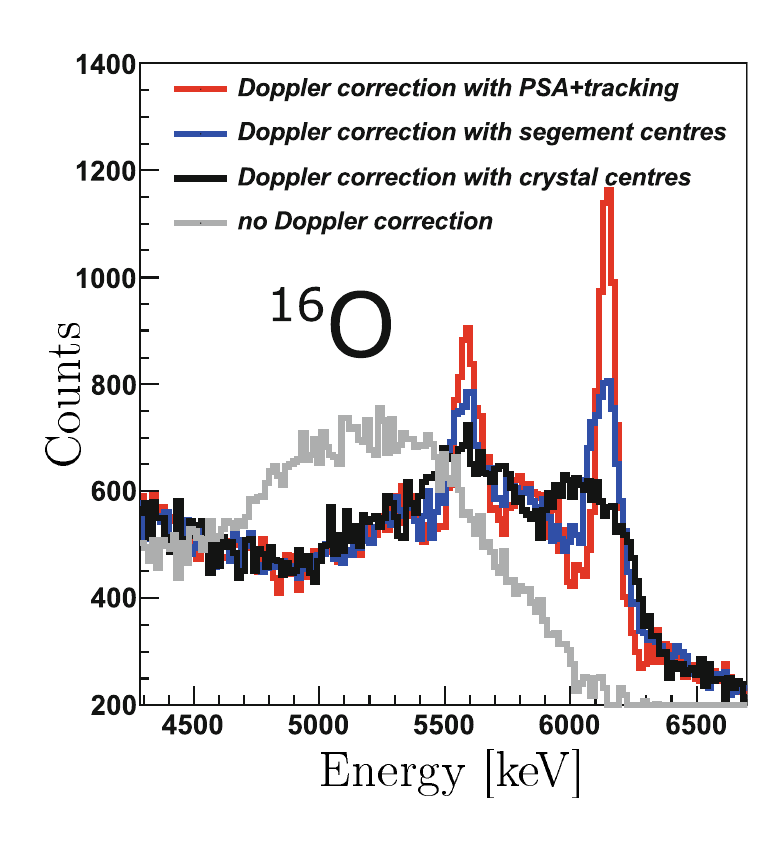}
  \end{center}
\caption{Energy spectrum of the $\gamma$ rays measured in
  correspondence of the (\nuc{17}{O},\nuc{16}{O}$^{\prime}$n$\gamma$)
  reaction channel~\cite{BRACCO2015}. The grey spectrum is without
  Doppler correction, while the others were corrected using position
  information at different precision level, as described in the
  legend. The case of a standard HPGe detector (i.e. with no position
  sensitivity) corresponds to the black line histogram.}
\label{pscstheo3}       
\end{figure}

As just mentioned, the \grt\ technology of AGATA allows reconstructing
the emission angle of the \gr\ with a precision of about 1 degree
and, consequently, to recover a large fraction of the energy
resolution degraded by the Doppler broadening. As a visual example,
the improvement in the quality of a \gdr\ spectrum obtained thanks to
\grt, can be appreciated by looking at the difference between the grey
line spectrum and its fully Doppler corrected version (red line) in
Fig.~\ref{pscstheo3}.

In order to appreciate in a more quantitative way the improvement in
resolution, it is important to consider in some detail the
experimental situation in which this spectrum was acquired. The theta
angle that appears in the Doppler shift formula is actually the angle
between the direction of the $\gamma$ and the velocity vector of the
emitting nucleus. Therefore, also the precision in the experimental
determination of the latter quantity is of relevance, in general, for
the Doppler correction quality. The 6.13 MeV $\gamma$ line, showed in
the figure, is emitted by the \nuc{16}{O} excited nucleus
(3$^-_1\rightarrow$0$^+_{g.s.}$) moving at a velocity of around the
20\% of the speed of light. This nucleus is produced in the reaction
(\nuc{17}{O},\nuc{16}{O}$^{\prime}$n$\gamma$), induced with a
\nuc{17}{O} beam at 20 MeV/u impinging on a \nuc{208}{Pb} target at
LNL lab~\cite{BRACCO2015}. This specific reaction channel was selected
by detecting the generated \nuc{16}{O} nuclei with a segmented Silicon
pad detector (TRACE~\cite{Gadea201188,MengPhd}) which allowed, in
fact, a precise experimental determination of their velocity vector.
This precise measurement of the velocity direction for the $\gamma$
emitting nucleus was crucial for allowing to maximise the quality of
Doppler correction and, also, to obtain \gdr\ angular distribution
plots that will be shown later.

The increase in resolving power from the combination of PSA and
\grt\ for prompt \gdr\ spectroscopy is clear when combined with
magnetic spectrometers giving the precise recoil vector of the
decaying nucleus. Lemasson et al. (see figure 3 in Lemasson et al. of
this issue) give as an example the spectroscopy of \nuc{98}{Zr} after
the fusion-fission with a \nuc{238}{U} beam at 6.2MeV/A and a
\nuc{9}{Be} target. Here the FWHM of the 1229.9 keV \gr\ from the
$2^{+}_{1}\rightarrow0^{+}_{1}$ transition varies from 15 keV for the
EXOGAM array to 5 keV for AGATA while covering similar angular ranges.

The good Doppler correction capability of AGATA makes it a powerful
tool for lifetime measurements using Doppler Shift methods such as
Recoil-Distance Doppler Shift (RDDS) or Doppler-Shift Attenuation
Method (DSAM). Using AGATA it is possible to measure lifetime from a
few fs with DSAM, to hundreds of ps with RDDS. Fast timing methods
using ancillary detectors can be used to complementary AGATA allowing,
with one experiment setup, the measurement of fs to ns. It is
therefore not surprising that in all three AGATA campaigns (Legnaro,
GSI, and GANIL) lifetime measurements have constituted a significant
fraction of the performed experiments. For a multitude of examples see
Bracco et al. \cite{BRACCO2021103887} and Lemasson et al.
\cite{Lemasson2023} and Gadea et al. \cite{Gadea2023} in this issue.

Detailed quantitative investigations have been performed in order to
study the impact of the position resolution from PSA in limiting the
Doppler correction quality in AGATA
(\cite{Soderstrom201196,RECCHIA2009555}). They all consistently point
to a $\approx$5 mm FWHM average position resolution value. This value,
however, can change depending on the region of the detector segment in
which the interaction took place and, of course, on the amount of
energy released in the $\gamma$ hit.

\FloatBarrier
\subsection{Background suppression}
\label{sec:ingrspthgrt:sec:bs}

In AGATA, the Compton background suppression is performed via
\grt\ (thus eliminating the necessity of using BGO shields). The
goodness of the suppression is quantified using the peak to total
ratio. The most recent work studying this parameter for AGATA
is~\cite{LJUNGVALL2020163297}. It was determined for $^{60}$Co source
data and compared with simulations. The peak-to-total value for the
1173 keV peak of $^{60}$Co was measured to be 36.4(4)\%. The
background is mainly due to single-interaction points considered,
erroneously, by the tracking as full-energy-peak events. Excluding
such events the peak-to-total is increased to 52.4(6)\%, but with a
reduction in efficiency of 17\%. In fact, the more stringent values of
tracking algorithm parameters we set, the cleaner spectra we obtain.
It is good practice to optimise the \grt\ parameters for each
experiment as the optimal values are \gdr\ energy and multiplicity
dependant.

Another kind of background that is commonly present in the experiments
is the one originating from neutrons. The possibility to suppress this
background is discussed in section~\ref{sec:ingrspthgrt:sec:ngd} of
this manuscript. Finally, since a properly tracked, and accepted,
\gr\ should always originate from the target position, the background
\grs\ originating from locations that are far from the target position
should be significantly reduced in the spectra. Different techniques
to achieve this are discussed in section~\ref{sec:griugrt} as they
fall under \gdr\ imaging techniques. This is particularly relevant for
experiments with relativistic exotic beams at GSI. In setups as
RISING~\cite{WOLLERSHEIM2005637}, in fact, sources of significant
background radiation were found to be materials of different origin
placed around the target (e.g. other detectors, degraders, the target
structure itself).

\FloatBarrier
\subsection{Angular distributions and Polarisation}
\label{sec:ingrspthgrt:sec:adap}

\begin{figure}[htp]
\begin{center}
  \includegraphics[width=\columnwidth,trim=0cm 6.2cm 0cm
    0cm,clip]{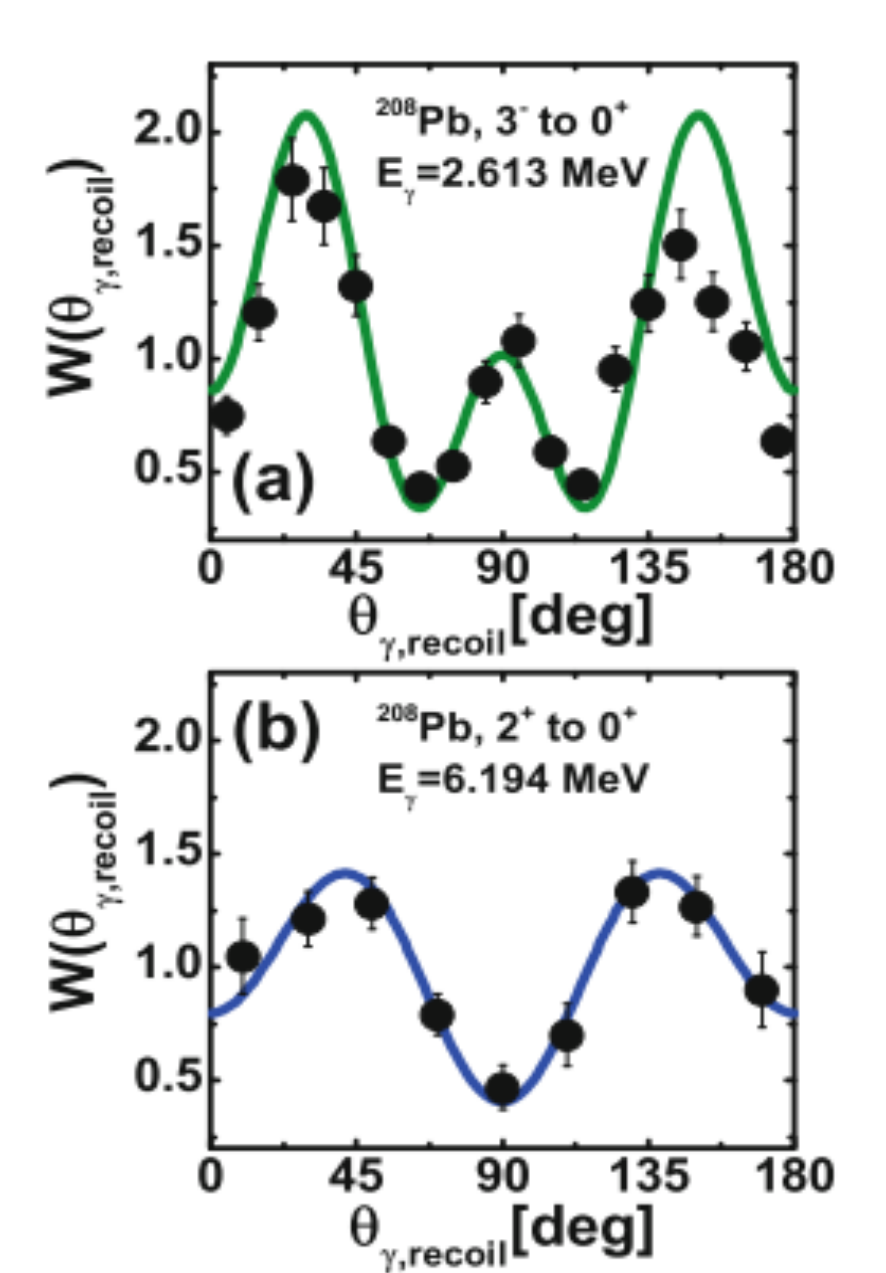}
  \end{center}
\caption{Distribution plots showing the angular correlation
  $\theta_{\gamma,recoil}$ (between the emitted $\gamma$ rays and
  velocity vector of the de-exciting recoiling nuclei), for two
  different transitions of $^{208}$Pb, each one having a distinct
  multipole character (octupole, quadrupole). Looking to Figure 10 in
  reference \cite{BRACCO2015} it can be appreciated how the angle
  between the vector associated to the $\gamma$ direction (included in
  the solid angle covered by the 5 ATCs) and the vector associated to
  the recoil direction (calculated on the basis of the \nuc{17}{O}
  angles covered by the solid angle of TRACE) spans from 0 to 180
  deg. The figure is adapted from reference \cite{BRACCO2015}.}
\label{pscstheo333}       
\end{figure}

The millimetric spatial sensitivity of AGATA detectors allows
measuring in a quasi-continuous way the emission direction of a
\gr\ emitted from the target position. AGATA is hence well suited for
measurements of angular distributions, angular correlations, and
\gdr\ polarisation. In many in-beam nuclear reactions the generated
degree of spin alignment allows observing the angular distribution of
\grs\ emitted following the de-excitation of a nucleus. This allows to
study the characteristic angular dependence due to the multipolarity
of the emitted \grs. In Fig.~\ref{pscstheo333} angular distributions
measured with AGATA are displayed. These results were extracted with
the experimental setup described in~\cite{BRACCO2015}. The angle
associated to the x-axis of the plots is the angle between the emitted
\gr\ and the velocity vector of the de-exciting recoiling nucleus (see
discussion of inelastic scattering reactions in
e.g.~\cite{PhysRevC.39.1307,SatchlerDR}). Measuring precisely the
direction of the recoiling nucleus, as was done in the experiment of
Bracco et al. \cite{BRACCO2015}, increases the alignment of the
reaction and gives very pronounced angular correlations. In the case
of the decay of nuclei that lack spin alignment or polarisation,
information on the multipolarity of the electromagnetic nuclear decays
can be extracted performing angular correlation measurements between
two \gdrs\ emitted in cascade. The use of AGATA for these kind of
measurements was investigated in~\cite{LJUNGVALL2020163297} using
source data.

From a physicist's point of view, it is appealing to have a device
capable of measuring with improved sensitivity the polarisation of
electromagnetic radiation. In AGATA, this is achieved thanks to the
PSA and tracking, providing an improved precision measurement of the
azimuthal angle in the Compton scattering process (i.e.
quasi-continuous angle Compton polarimetry \cite{ALIKHANI2012144}).
Regarding in-beam $\gamma$ spectroscopy for nuclear structure studies,
polarisation measurements are typically used for the determination of
the parity of nuclear excited states. Technical works were dedicated
to the investigation of the ability of AGATA detectors to measure the
polarisation of $\gamma$ rays \cite{TASHENOV2011164,Bizzeti2015}.
Finally, it is worth to mention that an additional technique, based on
a different principle than the quasi-continuous angle Compton
polarimetry (named Coulex-multipolarimetry with relativistic heavy-ion
beams), but still taking advantage of the position sensitivity of the
detectors, has been developed and
bench-marked~\cite{STAHL2015123,Napiralla2020}.

\FloatBarrier
\subsection{Neutron-$\gamma$ discrimination}
\label{sec:ingrspthgrt:sec:ngd}

It was early on recognised in the AGATA project that a $4\pi$ HPGe
detector array would detect neutrons emitted in nuclear reaction with
high efficiency. This can be viewed either as a problem - the neutrons
detected in AGATA generate background in the \gdr\ spectra, or as a
possibility to use AGATA as a neutron multiplicity filter. Neutron
induced background is not something unique to AGATA, and \lq\lq
neutron bumps\rq\rq\ associated with in-elastic scattering of neutrons
on germanium are a common feature in the \gdr\ spectra for
Compton-suppressed spectrometers such as EUROBALL~\cite{Simpson1997}
as well. The difference with a \grt\ array such as AGATA is the larger
solid-angle and that the \gdr\ tracking concept mixes energy
depositions from different detectors. This later aspect means that a
neutron interacting in one detector element can interfere with
\grs\ detected in adjacent detector elements.

Motivated by both the possibilities and potential issues with neutrons
in AGATA a set of investigations have been performed. The study by
Ljungvall et al.~\cite{LJUNGVALL2005553} sought to find pulse-shape
differences for signals originating from scattering neutrons or \grs.
Two different N-type HPGe detectors, one of closed-ended coaxial
geometry and one of planar geometry, were irradiated with neutrons and
\grs\ from a \nuc{252}{Cf} source. No significant difference between
signals originating from scattered neutrons or \grs\ was found. As the
possibility to separate interactions from neutrons and \grs\ based on
the pulse shapes is small, other methods have been developed. They aim
more at reducing the background generated by $(n,n'\gamma)$ reactions
on the Ge isotopes in the detectors rather than using AGATA as a
neutron multiplicity filter. In the work by Ljungvall et
al.~\cite{LJUNGVALL2005379} Geant4 Monte-Carlo simulations using the
AGATA code~\cite{Farnea2010331} were made for different neutron and
\gr\ distributions. The simulations show a detection efficiency for
neutrons from typical nuclear reactions of about 40\% for AGATA.
Furthermore, it was shown that counting the number of neutrons that
interacted in AGATA is challenging. AGATA can hence not be used as a
neutron multiplicity filter. This is easily understood, as the
majority of neutrons are detected by the \gr\ emitted in the inelastic
scattering process. Spatial and energy distributions of the
interaction of neutrons and \grs\ were also compared. The impact on
\grt\ performance was studied and quantified using two metrics,
Photo-peak efficiency $\epsilon_{ph}$, and for the Peak-To-Background
(PTB) defined as the area of the \gdr\ peak and the area of the
background in a region $\pm\sigma$ around the centroid of the
\gdr\ peak. The authors used a combination of forward tracking (OFT)
and back tracking \cite{LopezMartens2004454}. The authors concluded
that neutrons that scatter inside AGATA do reduce the efficiency and
PTB. However, they also pointed out that for traditional
\gdr\ spectrometers this effect is at least as large and hence not a
problem that is aggravated by \grt. This was an important result. It
should be noted that thanks to PSA it is possible to correct for hole
trapping coming from crystal damage from the fast neutrons. For
details see Boston et al. \cite{Boston2023} in this issue and
references therein. Ljungvall et al. \cite{LJUNGVALL2005379} also
investigated three different parameters to separate neutrons and
\grs\ from neutron interaction from \grs\ originating at the target
positions:
\begin{enumerate}
\item Time of Flight. For the nominal distance between AGATA and the
  target of 25 cm, the time resolution of the large volume HPGe
  crystals used in AGATA is however not sufficiently high.
\item $\Delta\left(\cos\theta\right)$, refers to the use of equation 3
  in figure~\ref{vertex} for the first vertex in a track. This is of
  course at the heart of \grt, but here the supplementary condition
  was only applied to the first vertex.
\item Ratio of low-energy interactions, defined as the number of
  interaction points with an energy below 20 keV to the total number
  of interaction points.
\end{enumerate}
The performance of the suggested methods were evaluated for
$\epsilon_{ph}$ PTB, and the $\Delta\left(\cos\theta\right)$ gives an
improved PTB of a factor of $3$ with a loss in $\epsilon_{ph}$ of
about 30\%.

A similar investigation, but using the MGT tracking
code\cite{bazzaccoMGT}, was performed by \atac\ et al.
\cite{ATAC2009554}. In addition to the condition on the difference
between the scattering angle of the first vertex calculated using the
interaction positions or the deposited energies they also investigated
the effects of gates on the energy of the first and second interaction
point in accepted tracks, and the use of a gate on the acceptance
limit. One can note that \atac\ et al. used a condition on the angle
and not the cosine of the angle, as was the case for Ljungvall et al.
There is also a noticeable difference in how the
$\Delta\left(\theta\right)$ (or $\Delta\left(\cos\theta\right)$)
condition is used. \atac\ et al. only applied it to clusters with more
than 2 interaction points as simulations had shown that accepted
clusters originating from neutron-induced \grs\ on average have one
interaction point more. The results show an improvement over the work
of Ljungvall et al.~\cite{LJUNGVALL2005379}, with a smaller loss in
Photo-peak efficiency.

\begin{figure}[tbh]
  \centering
  \begin{tikzpicture}
    \node[above right, inner sep=0] (image) at (0,0) {
      \includegraphics[width=.99\columnwidth,trim=11cm 10cm 1.5cm 12cm,clip]
                      {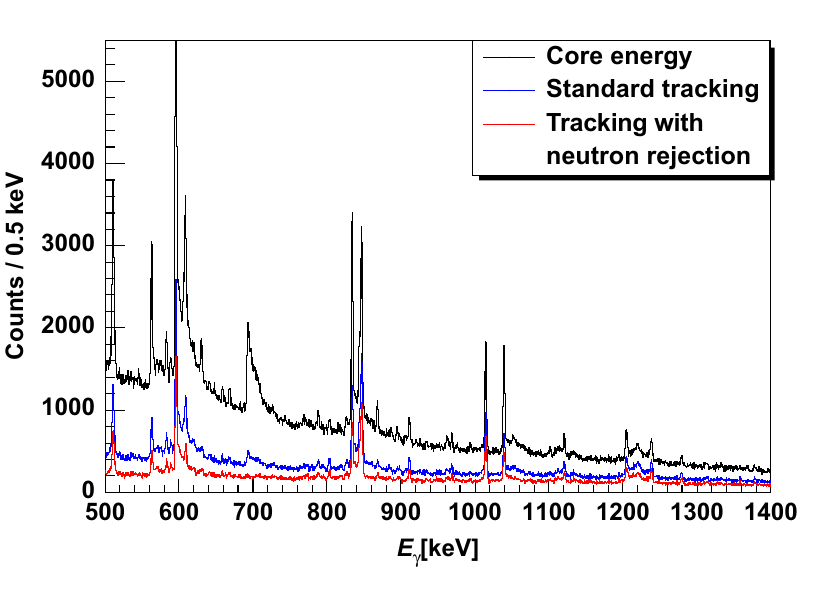}
    };
     \draw (1.9,2.5) ellipse (.25cm and 1.5cm);
  \end{tikzpicture}
                  \caption{Gamma-ray spectra showing the energy
                    deposition in AGATA when irradiated with neutrons
                    from a \nuc{252}{Cf} source. The black spectrum
                    corresponds to the sum of the core signals and
                    shows clear neutron-induced $\gamma$ rays with
                    corresponding triangular shapes. The blue line is
                    after \grt\ showing that already the standard
                    \grt\ discriminates well against neutron-induced
                    interactions. The red spectrum is after having
                    applied the additional gates used to suppress the
                    neutron-induced interaction. An ellipse highlights
                    the $\gamma$ transition $2^+_1\rightarrow 0^+_1$
                    in \nuc{74}{Ge} excited by (n,n'$\gamma$)
                    reactions. This figure is modified from
                    \senyigit\ et al.~\cite{SENYIGIT2014267}.}
                  \label{nhdgrs}
\end{figure}

The discrimination methods developed by \atac\ et al.
\cite{ATAC2009554} have also been applied to experimental data from
the demonstrator phase of AGATA. \senyigit\ et al.
\cite{SENYIGIT2014267} report on an experiment in which 4 AGATA triple
clusters were mounted together with 16 BaF$_{2}$ detectors from the
HELENA detector array. A \nuc{252}{Cf} source was positioned close to
the BaF$_{2}$ detectors and 50 cm from the AGATA detectors. The large
distance between the \nuc{252}{Cf} source and the AGATA detectors
allowed for a discrimination between neutron-induced interactions and
\gdr\ induced interactions using Time Of Flight. In figure
\ref{nhdgrs} \gr\ spectra from neutron induced reactions in the AGATA
detectors are displayed. Shown are the sum of core signals (in black),
the result of \grt\ (in blue), and tracked with conditions applied to
discriminate against neutron-induced \grs\ (in red). See the work of
\senyigit\ et al.~\cite{SENYIGIT2014267} for details. Already
\grt\ discriminates well against neutron-induced \grs, as the blue
spectrum contains much less intensity in the (n,n'$\gamma$) lines than
the black spectrum. An example of such a transition is marked with an
ellipse in figure \ref{nhdgrs}. This discrimination is easy to
understand as the \grs\ from (n,n'$\gamma$) do not originate from the
assumed target position. By applying gates on tracking parameters, see
\senyigit\ et al.~\cite{SENYIGIT2014267} a further reduction of
neutron-induced \grs\ is seen. The reduction of counts in the
\gdr\ spectra coming from (n,n'$\gamma$) is about 40\% (red spectrum).
Source data using \nuc{60}{Co} showed a loss of efficiency close to
20\% for \grs\ originating from the target position.

It has been proposed to look for isolated hits with an energy
of about 690 keV, corresponding to the E0 transition in \nuc{72}{Ge},
combined with PSA as a method to counts neutrons with
an efficiency of about 1.5\%~\cite{JENKINS2009457}. It is a too low
efficiency to be of interest for nuclear structure experiments, and is
too specialised for discrimination of the \gr\ background generated
by the fast neutrons. 

As a concluding remark on the work done to discriminate between
neutrons/neutron induced \grs\ and \grs\ one can state that the
developed methods do not allow the use AGATA as a neutron multiplicity
filter nor to completely remove the background induced by neutrons.
Their use on experimental data has, to our knowledge, up to now
been very limited.

\FloatBarrier
\section{Gamma-ray imaging techniques and $\gamma$-ray tracking}
\label{sec:griugrt}

\begin{figure}[hbt]
  \begin{center}
    \includegraphics[width=.99\columnwidth,trim=3.7cm 11cm 4cm 7.8cm,clip]
                      {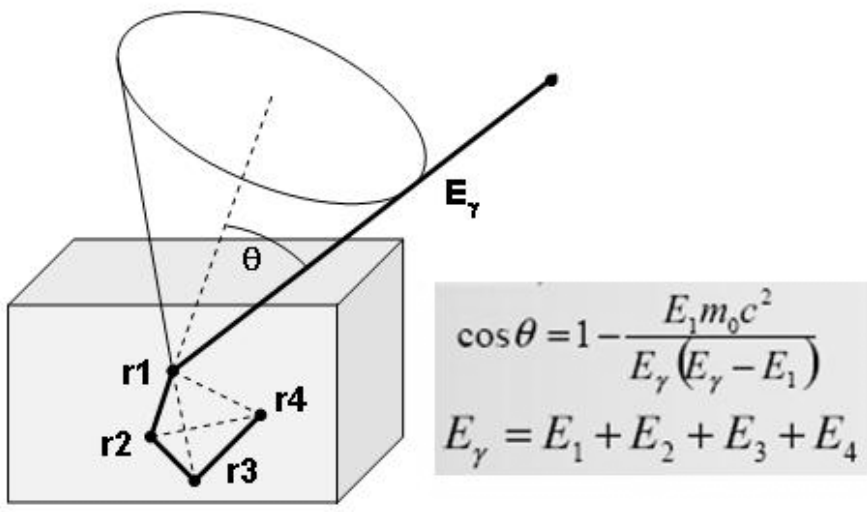}
                      \caption{Figure showing the principle of
                        \gr\ imaging using \grt. The crossing of many
                        cones builds up the image. Figure from J. Gerl
                       ~\cite{GERL2005688}.}
                      \label{gammarayimagingprinciple}
  \end{center}
\end{figure}

Early in the development of \grt\ the connection with \gdr\ imaging
was made: the ordering of the \gdr\ interaction allows the
reconstruction of a cone of origin for the \gr. This is illustrated in
figure~\ref{gammarayimagingprinciple}. An example of early exploratory
work (within what was to become the AGATA community) is that of van
der Marel et.~\cite{VANDERMAREL2001276}. The back tracking algorithm
was used to assess the performance of a setup consisting of two
(hypothetical) planer HPGe detectors. Rather promising results are
shown for SPECT and PET applications, although it should be mentioned
that very optimistic assumptions on the achievable position resolution
were made. Another study of possible applications using \grt, and this
time with AGATA crystals, is that of Gerl~\cite{GERL2005688}. It is
stated that the experimentally achieved position ($\sigma=2mm$) and
energy ($2.1$keV@$1.3$MeV FWHM) resolution is adequate for both
nuclear safety and medical applications.

Multiple investigations have also been made using AGATA detectors
combined with other detectors in order to improve the Compton imaging
performance~\cite{Moon_2011,Steinbach2017}. These investigations are
not in the scope of this article.

Compton imaging has also been used to characterise the performance of
AGATA~\cite{RECCHIA200960} and to improve the background rejection
using imaging techniques~\cite{DONCEL2010614}. In the work of Recchia
et al.\cite{RECCHIA200960} Compton imaging is used to estimate the
position resolution achieved by the PSA algorithms in
AGATA. In the paper the different contributions to the angular
resolution of \lq\lq cones\rq\rq\ (see figure
\ref{gammarayimagingprinciple}) given by the Compton formula are
calculated. It is shown that the angular resolution is dominated by
the incertitude coming from the PSA determining the
position of the \gdr\ interaction. Imaging can therefore estimate the
achieved position resolution. The method to estimate the position
resolution used by Recchia et al. is to compare images reconstructed
using experimental data and images reconstructed using simulations, in
which the assumed position resolution is varied. A \nuc{60}{Co} source
was placed 1 m from the AGATA detector. After PSA and a simplified
\grt, consisting in assuming that the interaction point with the
largest deposited energy is the first, the so-called back-projection
method~\cite{Wilderman1998} was used to create the images. By
comparing the FWHM of the projection on one axis of the image the
position resolution is deduced. The result is compatible with in-beam
measurements using the Doppler shift to deduce the PSA
resolution~\cite{Soderstrom201196}.

Another example of the connection between Compton imaging \grt\ is the
use of imaging to suppress background in \gr\ spectra. Doncel et al.
\cite{DONCEL2010614} used one AGATA detector to quantify how well
\grt\ can separate the origin of \grs. The setup consisted of the
AGATA detector surrounded by three \gdr\ source, \nuc{60}{Co},
\nuc{137}{Cs}, and \nuc{152}{Eu}.
\begin{figure}[hbt]
  \begin{center}
    \includegraphics[width=.55\textwidth,trim=4.25cm 2cm 2.cm
      8cm, clip] {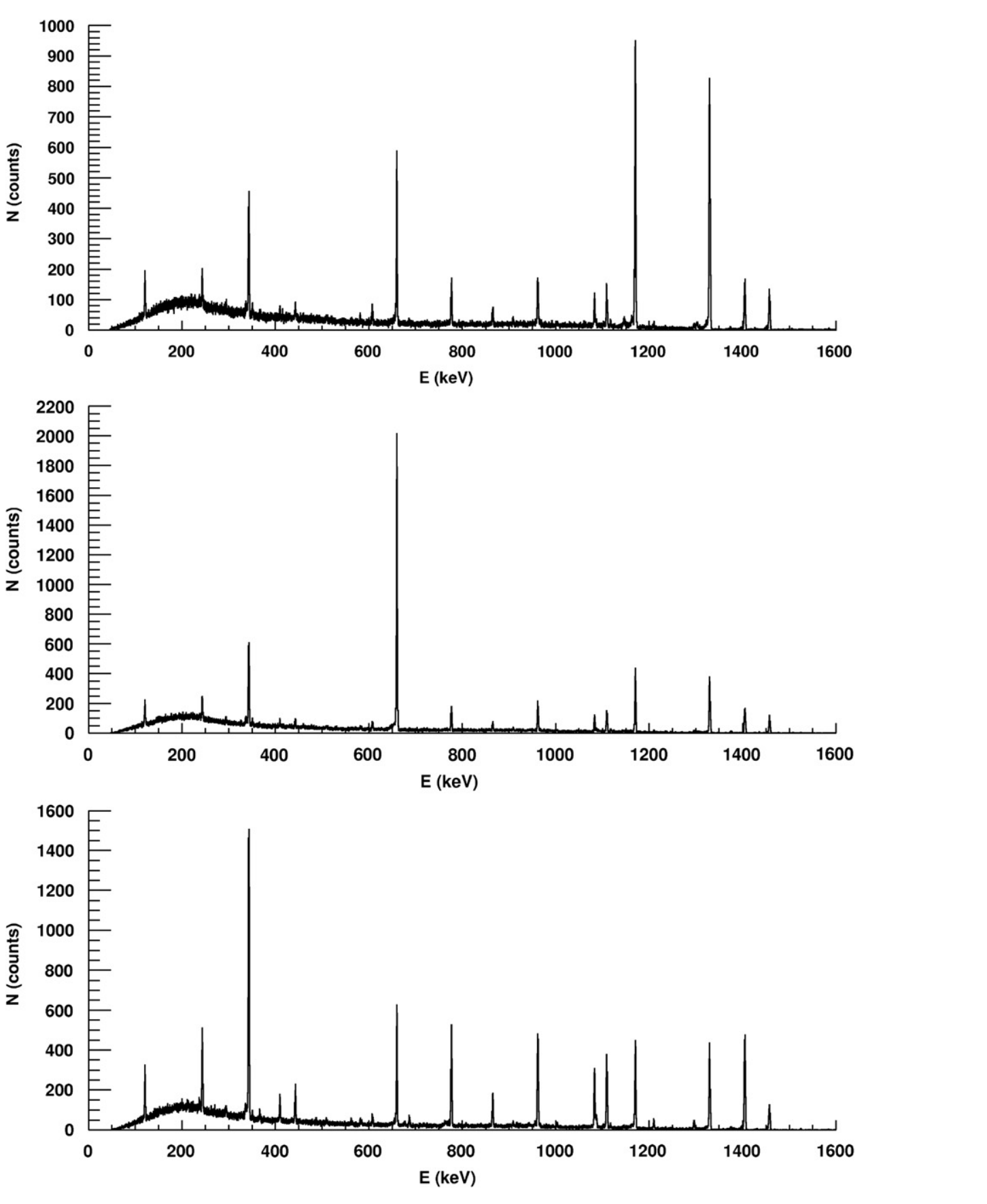}
                      \caption{Spectrum from \nuc{60}{Co}
                        incremented when the difference for
                        $\theta_{C}-\theta_{G}$ was smallest for the
                        position where the from \nuc{60}{Co} source
                        was located (top panel). Middle and bottom
                        panel show the same thing but for
                        \nuc{137}{Cs} and \nuc{152}{Eu}, respectively.
                        Figure from Doncel et al.\cite{DONCEL2010614}.}
                      \label{spectracomptonimagebkgsuppress}
  \end{center}
\end{figure}
Using a simplified tracking algorithm adapted to experimental setup
consisting of one symmetric AGATA crystal it was shown that comparing
the scattering angle as given by the energy deposition with the angle
coming from geometrical considerations a suppression of background
from \gdr\ sources of known origin with a factor of 3 is possible. In
figure~\ref{spectracomptonimagebkgsuppress} spectra incremented based
on the difference between the first scattering angle as calculated by
the first energy deposition $\theta_{C}$ and the angle given by the
source position, the first, and second interaction points, called
$\theta_{G}$. A clear enhancement for the \gr\ originating from the
correct source can be seen. This method is of interest to identify
unknown \grs\ in spectra by creating spectra assuming different origin
of the \grs\ when doing the \grt. To our knowledge this has however
not been exploited when performing \grt.

\FloatBarrier

\section{Conclusions and Perspective}

Over the last 15 years \grt\ has proved itself as a viable way of
constructing high-performance high-resolution \gdr\ spectrometers
using HPGe detectors. Developed using Geant4 simulations of ideal
$4\pi$ spheres the algorithms have been adapted to experimental data
and the defaults of PSA and provide performance close to what was
projected based on the simulations. The ability to identify the first
interaction point of a \gr\ with a precision of less than 5 mm is a
huge advantage for in-beam \gdr\ spectroscopy allowing to recover the
intrinsic high energy resolution of the HPGe detectors. It also allows
for high precision lifetime measurements using Doppler Shift methods
with an unprecedented sensitivity.

As this review is written it is believed, but not proven, that
improving the \grt\ algorithms presently used by the AGATA and GRETA
collaborations requires event-by-event errors on position (and energy
partitioning in the case of multiple hits in an segment) coming from
the PSA. A better estimation of the number of scatterings a
\gr\ really underwent before absorption is also welcome. A tracking
algorithm that possesses such complete information will provide better
discrimination between fully absorbed \grs\ and \grs\ that scattered
out of the array.

A closer coupling between PSA and \grt\ is also envisioned. This can
be either by a PSA providing multiple solution to \grt, and hence
relying on \grt\ to solve the problem of identifying the correct
number of interactions a \gr\ has undergone before absorption. Here
the general structure of the Data Acquisition system dividing AGATA
into local (detectors and hence PSA) and global levels (e.g. tracking)
can be kept. However, the combinatorial nature of the problem might
lead to computational challenges. An even more ambitious idea is to
combine PSA and \grt\ in one large minimisation process. It would not
only require an important computational effort but also exploratory
work investigating how to combine the figure-of-merits of PSA and
\grt.

The use of machine learning for \grt\ has already started and at the
time of writing this paper (2022) it seems to be a promising
avenue to follow.

\FloatBarrier

The authors would like to thank the AGATA collaboration. The AGATA
project is supported in France by the CNRS. Data used for this
publication were collected at INFN Legnaro, GSI, and GANIL and this
work would not have been possible without the valuable contributions
from these laboratories and their staff.

\bibliographystyle{elsarticle-numetal}
\bibliography{trackingbib}

\end{document}